\documentclass[fdp,fleqn]{w-art}

\usepackage{times}
\usepackage{w-thm}
\usepackage[]{graphicx}

\usepackage{color}

\newcommand \be{\begin{eqnarray}}
\newcommand \ee{\end{eqnarray}}

\usepackage{amsmath,amsfonts,amssymb}
\DeclareMathOperator{\Tr}{Tr}

\DeclareMathOperator{\Li}{Li}

\DeclareMathOperator{\sn}{sn}

\def\bR {\mathbb{R}}
\def\bS {\mathbb{S}}

\def \EE {{\mathbb{E}}}
\def \KK {{\mathbb{K}}}

\newcommand{\beq}{\begin{equation}}
\newcommand{\eeq}{\end{equation}}
\newcommand{\bal}{\begin{equation}\begin{aligned}}
\newcommand{\eal}{\end{aligned} \end{equation}}
\newcommand{\bea}{\begin{eqnarray}}
\newcommand{\eea}{\end{eqnarray}}

\newcommand{\vev}[1]{{\left< {#1} \right>}}

\newcommand{\eqn}[1]{(\ref{#1})}

\newcommand{\cN}{{\mathcal N}}
\newcommand{\cP}{{\mathcal P}}

\begin{document}
\DOIsuffix{theDOIsuffix}
\Volume{55}
\Issue{1}
\Month{01}
\Year{2007}

\vspace{ -3cm} \thispagestyle{empty} \vspace{-1cm}
\begin{flushright} ICCUB--12--004
\end{flushright}
\vspace{ -0.8cm}
\begin{flushright} kcl-mth--12--01
\end{flushright}
\pagespan{1}{6}
\Receiveddate{}
\Reviseddate{}
\Accepteddate{}
\Dateposted{}



\title[Short title]{Generalized quark-antiquark potential in AdS/CFT}


\author[F. Author]{Valentina Forini\inst{1,}%
  \footnote{Corresponding author\quad 
vforini@icc.ub.edu
       }}
\address[\inst{1}]{Institute of Cosmos Sciences and Estructura i Constituents de la Materia \\
Facultat de F\'isica, Universitat de Barcelona, Av. Diagonal 647, 08028 Barcelona, Spain}
\author[S. Author]{Nadav Drukker\inst{2,}\footnote{nadav.drukker@kcl.ac.uk}}
\address[\inst{2}]{Department of Mathematics, King's College London \\ The Strand, WC2R 2LS, London, UK}
\begin{abstract}
In this talk we present a family of Wilson loop operators which continuously interpolates between
the 1/2 BPS line and the antiparallel lines, and can be thought of as
calculating a generalization of the quark--antiquark potential for the gauge theory on $S^3\times\bf{R}$. 
We evaluate the first two orders of these loops perturbatively both in the gauge and string theory. We  obtain  analytical expressions in a systematic expansion around the 1/2 BPS configuration, and comment on possible all-loop patterns for these Wilson loops.

\end{abstract}
\maketitle                   



\renewcommand{\leftmark} {V. Forini and N. Drukker: Generalized quark-antiquark potential in AdS/CFT}



\section{Overview}

One of the most fundamental observables in a quantum field theory is the potential between charged particles, which in a gauge theory is captured by a long rectangular Wilson loop, or a pair of antiparallel lines representing the trajectories of infinitely heavy quarks.  Such quark-antiquark potential can be also considered in  the maximally supersymmetric  ${\cal N}=4$ SYM theory, where ``quarks'' are  modeled by infinitely massive W-bosons arising from a Higgs mechanism~\cite{maldacena-wl}.

The expectation value of this observable was calculated very early after the  introduction of the $AdS$/CFT correspondence by the effective action of a string ending along the curve on the four-dimensional $AdS$ boundary, and is in fact a seminal example of the duality itself.
In this context of a conformal field theory the potential is fixed to be Coulomb-like and the whole dynamical content is in the corresponding coefficient, for which the weak and strong coupling ('t Hooft coupling $\lambda$) previously obtained results read
 \be\label{c}
V_{q\bar q}\,(\lambda,L)=-\frac{1}{L}\,c(\lambda) ~,\quad\quad\quad c(\lambda)=\begin{cases} 
\frac{\lambda}{4\,\pi}\,\Big[1-\frac{\lambda}{2\,\pi^2}\Big(\ln\frac{2\pi}{\lambda}-\gamma_E+1\Big)+{\cal O}(\lambda^2)\Big],  & ~~~~~~~~~~~~\lambda\ll 1\\
\frac{\sqrt{\lambda}\,\pi}{4\,\KK(\frac{1}{2})^2}\, \,\Big[1+\frac{a_1}{\sqrt{\lambda}}+{\cal O}\Big(\frac{1}{(\sqrt{\lambda})^2}\Big)\Big], &~~~~~~~~~~~~ \lambda\gg 1\,.
\end{cases}
\ee
Above, $L$ is the distance between the lines, $\KK$ is the complete elliptic integral of the first kind and the weak-coupling expansion is the field-theoretical calculation of~\cite{essz,esz,Pineda}.  On the string theory (strong coupling) side, the question of evaluating the first quantum string correction  $a_1$ to the classical result of~\cite{maldacena-wl}~\footnote{This is actually the $AdS_5\times S^5$ counterpart of the so-called ``L\"uscher term'', which in flat space is a coulombic term proportional to the number of transverse dimensions~\cite{Luescher}.  
} is a hard mathematical problem.  
The absence of parameters in the problem (the only one, $L$, being fixed by conformal invariance) precludes considering special scaling limits in which nice results in $\sigma$-model perturbation theory have been obtained for some relevant string solutions (see, for example,~\cite{Roiban:2007dq,ftt} and reference therein). The  coefficient  $a_1$ was presented formally in \cite{fgt,dgt}, evaluated numerically in \cite{chr} to be $a_1=1.33459$ and simplified further in~\cite{vali-lines} to an analytic one--dimensional integral representation. 

It is hard to guess how to connect the two regimes of (\ref{c}). It is tempting to think about the chance of exploiting the integrability of the underlying AdS/CFT system and describe correctly the interpolation of $c(\lambda)$ between the two regimes of (\ref{c}), as  in the by now most famous example of smooth interpolation for a  non-protected quantity - the cusp anomaly of ${\cal N}=4$ SYM~\cite{cusp}.
\\

Our proposal~\cite{us} for addressing the problem relies on the introduction of \emph{extra parameters} in the initial setup. They do not make the perturbative or supergravity  calculation any harder and allow, in fact,  to interpolate between protected, much simpler, operators and  the desired observable.  
The first deformation parameter (indicated below with  $\theta$)  allows for the two lines to couple to two different scalar fields, and was already  introduced  in \cite{maldacena-wl}. In the general expression  of the Maldacena-Wilson loop 
\beq
W=\frac{1}{N}\Tr\cP\exp\left[\oint (iA_\mu\dot x^\mu+\Phi_I\Theta^I|\dot x|)ds\right],
\label{wl}
\eeq
we allow two different values of $\vec{\Theta}$ of relative angle $\theta$ on the two long edges of the rectangle. 
For $\theta=0$ the two lines couple to the same scalar 
field, say $\Phi_1$. When $\theta=\pi/2$ the two lines couple to 
$\Phi_1\pm\Phi_2$, which are orthogonal to each-other. 
Then for $\theta=\pi$ they couple to the field $\Phi_2$, but with opposite 
signs, which means that the lines are effectively parallel, rather than antiparallel. 
In that case the two lines share eight supercharges and the correlator is trivial.
The other deformation parameter  (indicated below with  $\phi$)  is geometric, and a way to illustrate it is to replace 
the theory on $\bR^4$ with the theory on $\bS^3\times\bR$ (related by the exponential 
map). We consider 
a pair of antiparallel lines separated by an angle $\pi-\phi$ on $\bS^3$. For 
$\phi=0$ the two lines are antipodal and mutually BPS, while for $\phi\to\pi$ the 
lines get very close together. ``Zooming in'' to the vicinity of the lines by a 
conformal transformation we get a situation very similar to the original antiparallel 
lines in flat space. An equivalent picture is that of a cusp in the plane in $\bR^4$. For $\phi=0$ 
the cusp disappears and the system is that of a single infinite straight line.

In the $\bS^3\times\bR$ picture the expectation value of the Wilson 
loop calculates the effective potential $V(\phi,\theta,\lambda)$ between a generalized 
quark-antiquark pair. 
%
In the case of a cusp in $\bR^4$ the loop suffers from logarithmic divergences \cite{Brandt}.  The expectation values of the loop in the two pictures are respectively 
\beq
\vev{W}\approx \exp\Big[{-}T\,V(\phi,\theta,\lambda)\Big]\,,\quad\quad\quad\quad \vev{W_\text{cusp}}\approx \exp\Big[{-}\log(R/\epsilon)\,V(\phi,\theta,\lambda)\Big]\,.
\label{cusp-vev}
\eeq
The logarithmic divergence is exactly the same as the linear time divergence, and the cutoffs of the two calculations are related by $\log(R/\epsilon)\sim T$.

The effective potential $V(\phi,\theta,\lambda)$ depends on the 't~Hooft coupling 
$\lambda=g^2N$ (we do not consider non-planar corrections) and it can be 
expanded at weak coupling and at strong coupling in the two relevant asymptotic  expansions
\be\label{V}
V(\phi,\theta,\lambda)=\begin{cases} 
\sum_{n=1}^\infty \left(\frac{\lambda}{16\pi^2}\right)^nV^{(n)}(\phi,\theta),  & ~~~~~~~~~~~~\lambda\ll 1\\
\frac{\sqrt\lambda}{4\pi}
\sum_{l=0}^\infty \left(\frac{4\pi}{\sqrt\lambda}\right)^lV_{AdS}^{(l)}(\phi,\theta), &~~~~~~~~~~~~ \lambda\gg 1~~.
\end{cases}
\ee
Below, we will present the evaluation of the first two terms of both regimes, adopting the picture of a cusp in $\bR^4$ at weak coupling and the $\bS^3\times \bR$ picture at strong coupling. In particular, at strong coupling the coefficients in the perturbative expansions are complicated functions of the 
angles $\phi$ and $\theta$ which are given only implicitly 
(at the classical level) or in integral form (one--loop). We consider therefore 
the expansion of these functions around $\phi=\theta=0$. This is an expansion 
around the $1/2$ BPS line (related to the circle via conformal transformation), one of the most simple observables in 
the theory. As a consequence, we obtain here \emph{analytic results} at both weak and strong coupling.  

Focussing on the first coefficients of this expansion, we argue below how they should receive contributions only from a subset of graphs in  perturbation theory -- the most connected graphs. At variance with the case of the circular Wilson loop, where in the Feynman gauge 
only ladder diagrams contribute and all interacting graphs combine to 
vanish \cite{esz, dg-mm,pestun}, we find here an observable 
which gets contributions only from the most interacting graphs. To our surprise, from the explicit calculation 
of the 2--loop graphs, we find that the result of these internally--connected graphs 
is simpler than the internally--disconnected one and does not involve polylogarithms.  
Since summing up ladder graphs is rather easy~\footnote{In~\cite{mos}, an integral equation was written whose solution gives the contribution
of ladder graphs to all orders in perturbation theory.}, it would be very interesting to explore the 3--loop graphs and see whether 
a similar pattern persists and perhaps learn how to calculate the most 
connected graphs to all orders.


In the rest of the talk we present a summary of our results at weak and strong coupling (Section 2), the explicit analytic expressions of the expansion around the BPS configuration and a short discussion on how the relevant coefficients can be evaluated via insertions of local operators into the loop (Section 3). The results obtained are suggestive of the framework in which an efficient description of the weak-to-strong coupling interpolation for these Wilson loops might take place. Certainly, they represent a set of analytic data to be of reference if an all-loop calculation will ever emerge.


\section{Results at weak and at strong coupling}
\label{sec:pert}
At \emph{weak coupling}, 
we work with the cusp in $\bR^4$~\cite{Polyakov:1980ca} and allow for an extra angle $\theta$ in $\cN=4$ SYM. 
For the potential $V(\phi,\theta)$ up to two-loops we found~\footnote{The calculation of $V^{(1)}$ at one--loop order was done in \cite{dgo}. The $\theta=0$ case is in \cite{mos} 
(see also \cite{kr-wl}), where expressions were written in integral form. Here we have extended the expressions  to $\theta\neq0$ and computed the integrals in closed form.} 
\bal
V^{(1)}(\phi,\theta)&=-2\,\frac{\cos\theta-\cos\phi}{\sin\phi}\,\phi\,;
\\
V^{(2)}(\phi,\theta)~
&=V^{(2)}_\text{lad}(\phi,\theta)+V^{(2)}_\text{int}(\phi,\theta)\,,
\\
V^{(2)}_\text{lad}(\phi,\theta)
&=-4\,\frac{(\cos\theta-\cos\phi)^2}{\sin^2\phi}
\left[\Li_3\left(e^{2i\phi}\right)-\zeta(3)
-i\phi\left(\Li_2\left(e^{2i\phi}\right)+\frac{\pi^2}{6}\right)
+\frac{i}{3}\phi^3\right],
\\
V^{(2)}_\text{int}(\phi,\theta)
&=\frac{4}{3}\,\frac{\cos\theta-\cos\phi}{\sin\phi}
\,(\pi-\phi)(\pi+\phi)\phi\,,
\label{1-2loops}
\eal
where $V^{(2)}$ is written as a sum of the contribution of ladder~\footnote{After subtracting the exponentiation of the $O(\lambda)$ term.} and interacting graphs. 

The analytic expressions (\ref{1-2loops}) undergo various checks. In the BPS case~\cite{zarembo},  where $\phi=\pm\theta$, then  $V^{(1)}=V^{(2)}=0$ as expected. At large imaginary angle, 
the prefactor of the linear term matches indeed  a quarter of the perturbative 
expansion of the cusp anomalous dimension~ \cite{Kotikov:2003fb}. 
Formulas (\ref{1-2loops}) also reproduce (and generalize) the antiparallel lines result of  \cite{essz}. Taking the $\phi\to\pi$ limit  and specializing to the case $\theta=0$, the resulting expression matches the one in  \cite{essz} with the  replacement $L\to\pi-\phi$. 
It is  interesting to notice that the complicated interacting graphs result in a contribution much 
simpler than the one due to the 2--loop ladder graph and without polylogarithmic functions~\footnote{Note the  uniform transcendentality three (when $e^{2i\phi}$ is considered rational) of  both interacting and ladder graphs at this order. }. 
Indeed it is proportional to the 1--loop result with a ratio which is just is a polynomial 
in $\phi$. 
\\

At \emph{strong coupling}, Wilson loops are described by macroscopic 
strings \cite{maldacena-wl, rey-yee}. 
The classical solutions are found in  global Lorentzian $AdS_5$~\footnote{This is the appropriate strong coupling dual of the gauge theory on $\bS^3\times\bR$. } starting from a time-independent ansatz,  the boundary conditions being lines separated  by $\pi-\phi$ on the boundary of AdS and $\theta$ on $S^5$.  The relevant solutions (written down in the case of $\theta=0$ in 
\cite{dgo} and for $\theta\neq0$ in Appendix C.2 of \cite{dgrt-big}) can be found for arbitrary values of $\phi$ and $\theta$
as the solutions of transcendental equations. The result for the generalized potential   is then found in terms of elliptic integrals $\KK$ and $\EE$~\footnote{The standard linear divergence for two lines along the boundary, canceled as usual by
a boundary term, is here removed.}
\beq
V^{(0)}_{AdS}(\phi,\theta)=\frac{\sqrt\lambda}{2\pi}\frac{2\sqrt{b^4+p^2}}{b\,p}
\left[\frac{(b^2+1)p^2}{b^4+p^2} \KK(k^2)-\EE(k^2)\right],
\label{classical-action}
\eeq
where the elliptic modulus $k$ and the parameter $b$ are functions of  $p, q$, which are in turn related   to $\phi,\theta$ via transcendental equations.

Quadratic fluctuations around the classical solution can be considered, based on the Nambu-Goto type action in the static gauge.
The mass matrix in the resulting quadratic fluctuation Lagrangian, depending in general on the two  parameters of the problem, becomes diagonal in the two limiting cases $\theta=0$ (equivalently $q=0$) and  $\phi=0$ (the limit $p\propto q\to\infty$). In particular, for these values  all the quadratic fluctuation operators, which have a trivial time dependence, can be written in the form  of one-dimensional single-gap \emph{Lam\'e} differential operators~\footnote{See also~\cite{Beccaria:2010ry}.}.
The latter point is crucial. It makes it possible to trade the explicit evaluation of the eigenvalue spectrum for the relevant operators with the resolution of the associated differential equation (an approach known as Gelfand-Yaglom method, see also the analysis in~\cite{us}). Relying on the knowledge of the solutions to the Lam\'e spectral problem, all fluctuations determinants can be then computed analytically. The resulting (regularized) effective action $\Gamma_\text{reg}$, which is the  ratio of determinants including the contribution of the trivial time direction  ${\cal T}\!\!=\!\!\int \!d\tau$, is then expressed as a single integral~\footnote{The integration variable $\omega$ in (\ref{gammareg}) is the Fourier-transformed $\tau$ variable $\partial \tau=-i\,\omega$.} and defines the one-loop correction to the generalized quark-antiquark potential as follows (e.g. in the $\theta=0$ case) 
\beq
\label{gammareg}
V^{(1)}_{AdS}(\phi,\theta)=\frac{\Gamma_\text{reg}}{T}=-\frac{{\cal T}}{2\,T}\lim_{\epsilon\to0}\int_{-\infty}^{+\infty}\frac{d\omega}{2\pi}
\ln\frac{\epsilon^2\omega^2\det^{8}{\cal O}_F^\epsilon}
{\det^{5}{\cal O}_0^\epsilon\det^2{\cal O}_1^\epsilon\det{\cal O}_2^\epsilon}\,.
\eeq  
The explicit expressions for the 1d determinants can be found in~\cite{us}, here  we report as representative the bosonic contribution
\begin{equation}
\label{O2epsilon}
\det{\cal O}_2^\epsilon
\cong-\frac{\sinh(2\KK(k_2^2)\,Z(\alpha_2))}
{\epsilon^2\,\omega\sqrt{\omega^4+(2-4k^2)\omega^2+1}}\,,\quad\quad\quad\sn(\alpha_2|k_2^2)=\textstyle{\frac{\sqrt{1+k_2^2+\omega_2^2}}{k_2}}\,,
\end{equation}
where $Z$ is the Jacobi Zeta function,  $\sn$ is the Jacobi elliptic sine, $k_2$ is a rational function of $k$ and  $\omega_2$ a rational function of  $k$ and $\omega$.
Above,  $\epsilon$ is the standard infrared regulator curing the linear divergence expected at the boundary, the determinant is taken at leading order in a $\epsilon\simeq0$ expansion and an explicit subtraction of the remaining divergences (a regularization artifact) is made.

It is possible to see that both the classical and the one-loop strong coupling results, (\ref{classical-action}) and (\ref{gammareg})-(\ref{O2epsilon}), reproduce the known expressions for the antiparallel lines, in~\cite{maldacena-wl,rey-yee} and~\cite{chr,vali-lines} respectively, in the $\phi\to\pi, ~\theta=0$ limit~\footnote{This limit  translate in the conditions $p\to0\,,~ \frac{q^2}{p}=\text{fixed}, ~k^2=1/2$  on the parameters relevant at strong coupling.}. This happens, as in the weak coupling case, once the replacement of the pole $\pi-\phi\to L$ is performed.

It  is straightforward to evaluate the integral  (\ref{gammareg})  numerically for arbitrary values of   $\phi$, as well as in the analog case of $\phi=0$ and arbitrary $\theta$, while,  in general, we do not know how to calculate it analytically~\footnote{See however the results of~\cite{vali-lines} in the limit of antiparallel lines.}. 
To gain more analytic control over the form of $V_{AdS}^{(1)}$ we will  proceed in a 
 systematic expansion around $\theta=0$ and $\phi=0$, to which the next section is devoted.

\section{Near straight-line expansion}

In the  $\phi\to0$ limit  the cusp disappears and we are left with an infinite straight line in $\bR^4$, or a pair of antipodal lines on $\bS^3\times\bR$. In this case the analysis indeed simplifies, and allows for explicit analytic expressions at weak and at strong coupling.


At \emph{weak coupling}, the first few orders in the expansion of  \eqn{1-2loops} around $\phi=\theta=0$ read
\bal\label{Expweak}
V^{(1)}(\phi,\theta)=&\,
\theta^2-\phi^2
-\frac{1}{12}(\theta^2-\phi^2)^2+O((\phi,\theta)^6)\,,\\
V^{(2)}(\phi,\theta)=&\,-\frac{2\pi^2}{3}(\theta^2-\phi^2)
+\frac{1}{18}(\pi^2(\theta^2-\phi^2)^2+6(\theta^2-\phi^2)(3\theta^2-\phi^2))+O((\phi,\theta)^6)\,.
\eal
All the terms are proportional to $\theta^2-\phi^2$, and indeed we expect $V(\phi, \theta,\lambda)$ 
to vanish for $\theta=\pm\phi$, which are BPS configurations~\cite{dgrt-big}.

At \emph{strong coupling}, an expansion of the leading semiclassical result leads to
\bal
\label{V0ads-expand}
V_{AdS}^{(0)}(\phi,\theta)
=\,
\frac{1}{\pi}(\theta^2-\phi^2)
-\frac{1}{8\pi^3}(\theta^2-\phi^2)\left(\theta^2-5\phi^2\right)+O((\phi,\theta)^{6})\,.
\eal
At one--loop order in $\sigma$-model perturbation theory, the expansion translates in a small $k$ expansion of 
all the elliptic functions in the integrand of (\ref{gammareg}), and results in a power series of regular hyperbolic functions. An integration over the logarithm of this series can then always be performed, and gives 
\bal
\label{V1ads-expand}
V_{AdS}^{(1)}(\phi,0)
=&\,
\frac{3}{2}\frac{\phi^2}{4\pi^2}
+\left(\frac{53}{8}-3\,\zeta(3)\right)\frac{\phi^4}{16\pi^4}
+\left(\frac{223}{8}-\frac{15}{2}\zeta(3)-\frac{15}{2}\zeta(5)\right)\frac{\phi^6}{64\pi^6}+O(\phi^{8})\,.
\\
V_{AdS}^{(1)}(0,\theta)
=&\,
{-}\frac{3}{2}\frac{\theta^2}{4\pi^2}
+\left(\frac{5}{8}-3\,\zeta(3)\right)\frac{\theta^4}{16\pi^4}
+\left(\frac{1}{8}+\frac{3}{2}\zeta(3)-\frac{15}{2}\zeta(5)\right)\frac{\theta^6}{64\pi^6}+O(\theta^{8})\,.
\eal

Focus now on the expansion coefficients around $\phi=\theta=0$, for example  the first (quadratic) one  
\beq
\label{theta^2}
\frac{1}{2}\frac{\partial^2}{\partial\theta^2}V(\phi,\theta,\lambda)\Big|_{\phi=\theta=0}=
-\frac{1}{2}\frac{\partial^2}{\partial\phi^2}V(\phi,\theta,\lambda)\Big|_{\phi=\theta=0}=
\begin{cases}
\displaystyle
\frac{\lambda}{16\pi^2}-\frac{\lambda^2}{384\pi^2}+\cdots
\quad& \lambda\ll1\,,\\[4mm]
\displaystyle
\frac{\sqrt{\lambda}}{4\pi^2}-\frac{3}{8\pi^2}+\cdots
\quad& \lambda\gg1\,.
\end{cases}
\eeq
The  expansion around the 1/2 BPS  straight line can be viewed as a \emph{deformation} of the straight line  itself, and as such it can be written in terms of insertions of local operators into the Wilson loop. 
One can write the latter as a straight ($\phi=0$) line in the $x^1$ direction with arbitrary $\theta$
\beq
W=\frac{1}{N}\Tr\cP\left[\exp\Big(\int_{-\infty}^0 (iA_1+\Phi_1)ds\Big)
\exp\Big(\int_0^\infty (iA_1+\Phi_1\cos\theta+\Phi_2\sin\theta)ds\Big)\right]~,
\eeq
such that it couples to the scalar  $\Phi_1$ for all $s<0$ and to the linear combination  $\Phi_1\cos\theta+\Phi_2\sin\theta$ for $s>0$~\footnote{We fixed the parameterization such that $|\dot x|=1$, so we can ignore the difference between $x^\mu(s_i)$ and $s_i$.  }. 
Using that~\footnote{The first identity is the definition of $V$. The second follows from $\frac{\partial}{\partial\theta}\vev{W}=0$ and from  $\vev{W|_{\phi=\theta=0}}=1$.}~\beq
\frac{\partial^2}{\partial\theta^2}V(0,0)=
-\frac{1}{\ln(R/\epsilon)}\frac{\partial^2}{\partial\theta^2}\log\vev{W}\approx
-\frac{1}{\ln(R/\epsilon)}\frac{\partial^2}{\partial\theta^2}\vev{W},
\eeq
one finds for the coefficient in (\ref{theta^2})~\footnote{The variation with respect to $\theta$ is somewhat simpler than the the variation with respect to $\phi$, since the latter modifies the path of the loop and is captured by insertions of the field
strength $F_{\mu\nu}$ as well as its derivatives into the loop, while the former only introduces local scalar field insertions.}
\bal
\frac{1}{2}\frac{\partial^2}{\partial\theta^2}V=
&\,-\frac{1}{\ln(R/\epsilon)}\frac{1}{2N}\int_0^\infty ds_1\int_0^\infty ds_2\,
\vev{\Tr\cP\left[\Phi_2(s_1)\Phi_2(s_2)\,e^{\int_{-\infty}^\infty (iA_1+\Phi_1)ds}\right]}
\\&\,
+\frac{1}{\ln(R/\epsilon)}\frac{1}{2N}\int_0^\infty ds_1\,
\vev{\Tr\cP\left[\Phi_1(s_1)\,e^{\int_{-\infty}^\infty (iA_1+\Phi_1)ds}\right]}.
\label{insertions}
\eal
Examining the right-hand side is suggestive of a pattern expected to hold for all values of the coupling. One notices that graphs which involve propagators between the Wilson loop and itself, and not the insertions,  will vanish  due to the BPS nature of the straight line. At one and two-loop order, only graphs with at most one internally connected component contribute, as the  explicit 
expansion of $V^{(2)}_\text{int}$ and $V^{(2)}_\text{lad}$ in \eqn{1-2loops} easily confirms. The interesting observation is that this  argument should apply also to higher order graphs. \emph{Only graphs with one set of connected internal 
lines attached to the Wilson loop  contribute to this term}%
\footnote{This statement is true assuming the cancellation for the straight line 
does not require integration. Otherwise, there will be boundary terms in the 
disconnected graphs, which can be regarded as connected ones.}. Regarding further expansion coefficients, the one of $\theta^4$ will involve for example graphs with at most two disconnected internal  components, and so on. 
Since by explicit calculation we found that the connected (interacting) 
graphs at 2--loop order had a simpler (without polylogarithms) functional form than the disconnected (ladder) 
ones, it would be certainly interesting to see if this structure persists at 
higher orders in perturbation theory and whether it is possible to guess the answer 
for the most connected graphs at all loop order, and reproduce the strong 
coupling results in \eqn{theta^2}.

\begin{acknowledgement}

We thank the organizers of the XVII European Workshop on String Theory in Padua for creating an enjoyable and stimulating meeting.
 
\end{acknowledgement}


\begin{thebibliography}{10}


\bibitem{maldacena-wl}
J.~M.~Maldacena,
  Phys.\ Rev.\ Lett.\  {\bf 80}, 4859 (1998)
  [arXiv:hep-th/9803002].

\bibitem{essz}
  J.~K.~Erickson, G.~W.~Semenoff, R.~J.~Szabo and K.~Zarembo,
  Phys.\ Rev.\ D {\bf 61}, 105006 (2000)
  [hep-th/9911088].
  
  
\bibitem{esz}
  J.~K.~Erickson, G.~W.~Semenoff and K.~Zarembo,
  Nucl.\ Phys.\  B {\bf 582}, 155 (2000)
  [arXiv:hep-th/0003055].

  
\bibitem{Pineda}
  A.~Pineda,
  Phys.\ Rev.\  D {\bf 77}, 021701 (2008)
  [arXiv:0709.2876 [hep-th]].
  
 
\bibitem{Luescher} 
  M.~Luscher,
  Nucl.\ Phys.\ B {\bf 180}, 317 (1981);
   M.~Luscher, K.~Symanzik and P.~Weisz,
  Nucl.\ Phys.\ B {\bf 173}, 365 (1980).
  
  \bibitem{Roiban:2007dq}
  R.~Roiban and A.~A.~Tseytlin,
  JHEP {\bf 0711}, 016 (2007)
  [arXiv:0709.0681 [hep-th]].


 
\bibitem{ftt}
S.~Frolov, A.~Tirziu and A.~A.~Tseytlin,
  Nucl.\ Phys.\  B {\bf 766}, 232 (2007)
  [arXiv:hep-th/0611269].

 \bibitem{fgt}
  S.~Forste, D.~Ghoshal and S.~Theisen,
  JHEP {\bf 9908}, 013 (1999)
  [arXiv:hep-th/9903042].

\bibitem{dgt}
  N.~Drukker, D.~J.~Gross and A.~A.~Tseytlin,
  JHEP {\bf 0004}, 021 (2000)
  [arXiv:hep-th/0001204].
  
  
\bibitem{chr}
 S.~x.~Chu, D.~Hou and H.~c.~Ren,
  JHEP {\bf 0908}, 004 (2009)
  [arXiv:0905.1874 [hep-ph]].

\bibitem{vali-lines}
  V.~Forini,
  JHEP {\bf 1011}, 079 (2010)
  [arXiv:1009.3939 [hep-th]].

\bibitem{cusp} 
 N.~Beisert, B.~Eden and M.~Staudacher,
  J.\ Stat.\ Mech.\  {\bf 0701}, P01021 (2007)
  [hep-th/0610251];
  B.~Basso, G.~P.~Korchemsky and J.~Kotanski,
  Phys.\ Rev.\ Lett.\  {\bf 100}, 091601 (2008)
  [arXiv:0708.3933 [hep-th]].


\bibitem{us}
  N.~Drukker and V.~Forini,
  JHEP {\bf 1106}, 131 (2011)
  [arXiv:1105.5144 [hep-th]].
  

\bibitem{Brandt} 
  R.~A.~Brandt, F.~Neri and M.~-A.~Sato,
  Phys.\ Rev.\ D {\bf 24}, 879 (1981); 

  \bibitem{dg-mm}
  N.~Drukker and D.~J.~Gross,
  J.\ Math.\ Phys.\  {\bf 42} (2001) 2896
  [hep-th/0010274].
  
 
 
\bibitem{pestun} 
  V.~Pestun,
  arXiv:0712.2824 [hep-th].


\bibitem{mos}
  Y.~Makeenko, P.~Olesen and G.~W.~Semenoff,
  Nucl.\ Phys.\ B {\bf 748}, 170 (2006)
  [hep-th/0602100].

\bibitem{Polyakov:1980ca} 
A.~M.~Polyakov,
  Nucl.\ Phys.\ B {\bf 164}, 171 (1980).

\bibitem{dgo}
 N.~Drukker, D.~J.~Gross and H.~Ooguri,
  Phys.\ Rev.\  D {\bf 60}, 125006 (1999)
  [arXiv:hep-th/9904191].



\bibitem{kr-wl}
  G.~P.~Korchemsky and A.~V.~Radyushkin,
  Nucl.\ Phys.\ B {\bf 283}, 342 (1987).

\bibitem{zarembo}  
K.~Zarembo,
  Nucl.\ Phys.\  B {\bf 643}, 157 (2002)
  [arXiv:hep-th/0205160].


\bibitem{Kotikov:2003fb}
  A.~V.~Kotikov, L.~N.~Lipatov and V.~N.~Velizhanin,
  Phys.\ Lett.\ B {\bf 557}, 114 (2003)
  [hep-ph/0301021].
  
  
\bibitem{rey-yee} 
  S.~-J.~Rey and J.~-T.~Yee,
  Eur.\ Phys.\ J.\ C {\bf 22}, 379 (2001)
  [hep-th/9803001].
  
  

\bibitem{dgrt-big}
  N.~Drukker, S.~Giombi, R.~Ricci and D.~Trancanelli,
  JHEP {\bf 0805}, 017 (2008)
  [arXiv:0711.3226 [hep-th]].

\bibitem{Beccaria:2010ry} 
  M.~Beccaria, G.~V.~Dunne, V.~Forini, M.~Pawellek and A.~A.~Tseytlin,
  J.\ Phys.\ A {\bf 43}, 165402 (2010)
  [arXiv:1001.4018 [hep-th]].
  



\end{thebibliography}
\end{document}